\documentclass[aps,10pt,twocolumn]{revtex4}
\usepackage{graphicx}
\usepackage{amssymb}
\usepackage{amsfonts}
\usepackage{amsmath}
\usepackage{graphicx}
\usepackage{times}
\usepackage{dcolumn}
\usepackage{bm}
\usepackage{revsymb}
\usepackage{color}%
\usepackage{epstopdf}
\usepackage{mathrsfs}

\begin{document}

%\title{Should EVs try to coast or employ regenerative braking? \\
%\small Some physics considerations on how to brake in an electric vehicle}
%\title{The physics of optimal braking in an electric vehicle}
\title{Optimizing regenerative braking - a variational calculus approach}
\author{L.Q. English$^1$, A. Mareno$^2$, and Xuan-Lin Chen$^3$}
\affiliation{$^{1}$Department of Physics, Dickinson College, Carlisle, Pennsylvania 17013, USA}
\affiliation{$^{2}$Department of Mathematics and Computer Science, Pennsylvania State University, Capital College, Middletown PA, 17057, USA}
\affiliation{$^{3}$Department of Fluid Physics, Pattern Formation, and Biocomplexity, Max Planck Institute for Dynamics and Self-Organization, 37077 G\"{o}ttingen, Germany}

\begin{abstract}
We examine some basic physics surrounding regenerative braking and air drag. First, we analyze under what conditions it becomes energetically favorable to use aggressive regenerative braking to reach a lower speed over ``coasting'' where one relies solely on air drag to slow down. We then proceed to reformulate the question as an optimization problem to find the velocity-profile that maximizes battery charge. Making a simplifying assumption on battery-charging efficiency, we formulate the recovered energy as an integral quantity and solve the associated Euler-Lagrange equation to find the optimal braking curve. Using Lagrange-multipliers, we also explore the effect of adding a fixed-displacement constraint.  
\end{abstract}

\maketitle

\section{Introduction}
Regenerative braking is a process by which some of the initial kinetic energy of a vehicle, rather than wasted as heat, is instead recovered by converting it into electrical energy which in turn is stored as chemical energy in the battery \cite{clegg}. In the engineering literature, this topic is usually approached from a technological perspective \cite{eng1,eng2,eng3,eng4,eng5}. In contrast, here we take a more simplified but fundamental ``classical-mechanics'' and variational calculus approach to the subject. 

There has recently been some debate over the benefits of ``1-pedal driving'' - this is a mode where as soon as the driver of an electric vehicle (EV) releases pressure on the accelerator pedal, the electric generator is engaged, the battery charges up, and the EV begins to slow down. Tesla, for instance, makes this mode standard in all its cars. Other manufacturers, such as VW, have opted to preserve the driving experience of conventional internal-combustion-engine (ICE) cars, whereby releasing the accelerator pedal allows the car to coast. Coasting means that neither the motor nor the generator are engaged, and the vehicle slows down solely due to aerodynamic drag and other friction. Some people have argued that coasting is, in fact, always best in terms of efficiency, assuming no constraints on the stopping distance \cite{cleantechnica}. 

In this paper, we start by comparing the two strategies over the same fixed displacement, and we find that aggressive regenerative braking is superior to coasting as long as the final speed is small enough. While it is true that both the electric motor and the generator do not operate at perfect efficiency and produce losses, lowering the speed faster lowers the drag force on the vehicle (relative to coasting). Thus, there exist a trade-off: regenerative braking recovers some of the initial kinetic energy (though not all of it), but the electric motor then has to use additional energy to keep the vehicle in motion. Is more energy initially stored in the battery as is later drawn from it? As we will see, the answer often is yes.

In the second part of the paper, we reformulate the question in terms of an optimization problem. Here we ask: what should the speed-profile be that leaves the battery in the largest state of charge (SOC)? This recovered energy is an integral quantity, and we thus formulate the associated Euler-Lagrange equation whose solution maximize this quantity and yields the optimal velocity as a function of time. Modeling the regenerative-braking efficiency in the simplest, most tractable way, we solve this nonlinear differential equation numerically, as well as analytically in integral form. Finally, we examine the effect of adding a fixed-displacement constraint via a Lagrange multiplier. 

\section{Recoverable Energy Considerations}\label{sec1}
\subsection{Coasting}
In this scenario we would like to rely solely on air drag and rolling friction to slow our EV down from an initial speed of $v_i$ to a final speed of $v_f$.  
Here we ignore the contributions of rolling friction - it only exhibits a weak dependence on vehicle speed \cite{katz} and should thus result in a similar energy loss when comparing coasting with regenerative braking. What is the final state of the battery relative to its initial state while coasting? It is clear that they are the same. No energy is transferred to or from the battery.

For the sake of comparing to the regenerative-braking scenario, let us calculate the accumulated displacement for this coasting process. Let's assume that the drag force is quadratic in velocity, such that,
\begin{equation}
F_D=\frac{1}{2}\rho A C_d v^2 = D v^2,
\end{equation}
where $A$ is the cross-sectional area, $\rho$ the density of air, and $C_d$ the drag coefficient, and where we have combined these parameters into a single constant, $D$. This form of the drag force is valid for relatively high speeds, such as a car on a highway, but fails at low speeds. We thus arrive at the following differential equation governing coasting,
\begin{equation}
\label{coast}
m\dot{v} = -D v^2,
\end{equation}  
where $m$ is the car mass. Equation (\ref{coast}) has the well-known analytical solution,
\begin{equation}
\frac{1}{v(t)}=\frac{1}{v_i}+\left(\frac{D}{m}\right) t.
\label{tc}
\end{equation}
The displacement, $\Delta x = \int_0^{t_c} v(t) dt$ can then be found, after a few steps, 
\begin{equation}
\label{displ}
\Delta x = \frac{m}{D} \ln \left(\frac{v_i}{v_f}\right).
\end{equation}
It is clear from Eq.(\ref{displ}) that we cannot use the quadratic-drag assumption down to a $v_f$ of zero.

\subsection{Regenerative Braking}
\begin{figure}
\includegraphics[width=0.5\textwidth]{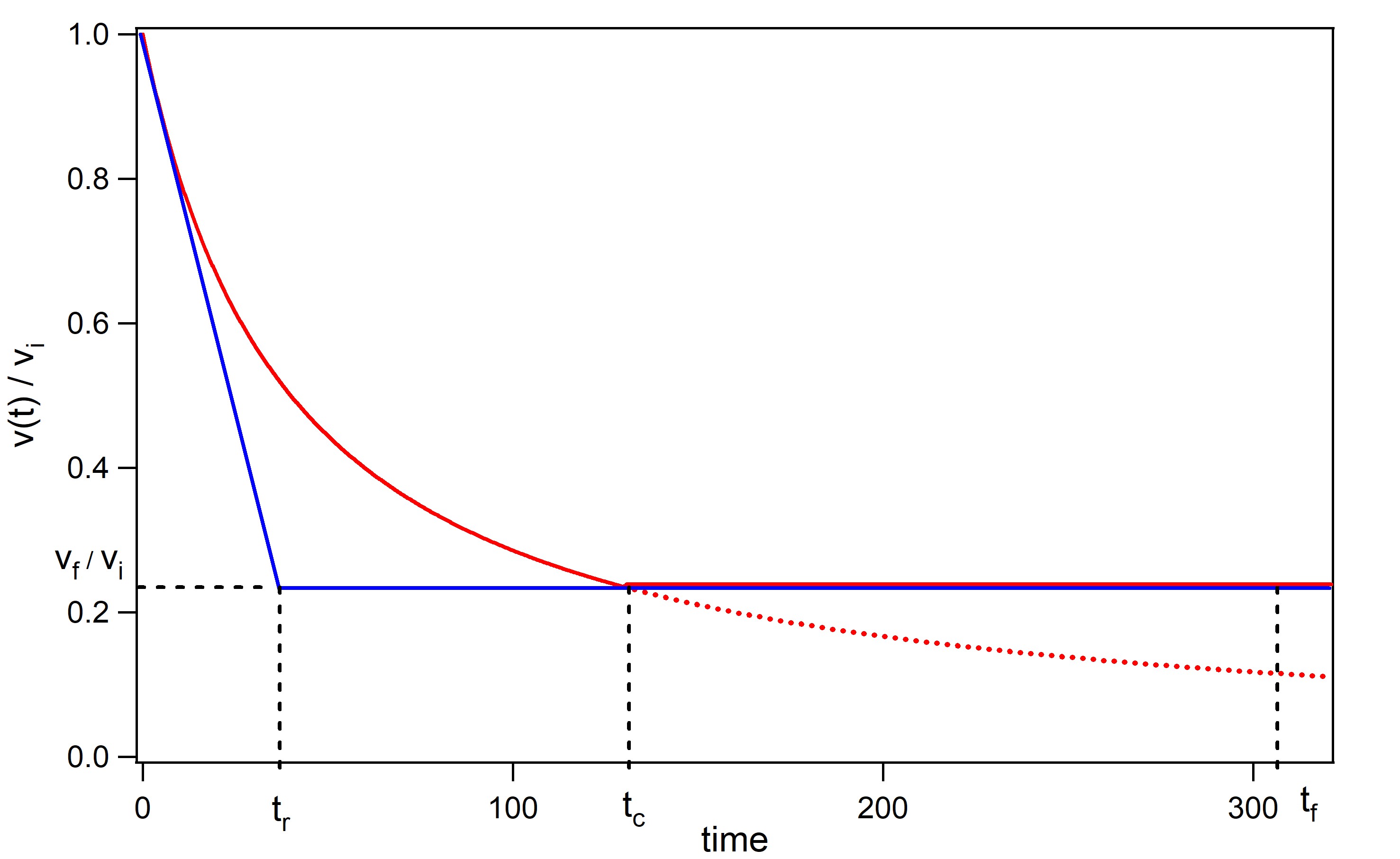}
\caption{Speed as a function of time for the two processes: coasting (red line) and regenerative braking (blue line) from $v_i$ to $v_f$. The relevant times are indicated on the horizontal axis.}
\label{fig1}
\end{figure}

Instead of coasting, let us now try a different strategy: we will slow the EV down quickly from $v_i$ to $v_f$ using regenerative braking, and then we will continue to drive at $v_f$ until we have covered the same distance, $\Delta x$, as during coasting before. For sake of simplicity, let us further assume that in the braking phase, the speed drops linearly in time, and that we accomplish this part in a time, $t_r$, as shown in Fig.\ref{fig1}. As we will see in Section \ref{optimal}, this constant-acceleration strategy is not energy optimal and can still be improved upon, but it serves as a good starting point. The total displacement can now be expressed as,
\begin{equation}
\Delta x = v_f t_r + \frac{1}{2} (v_i-v_f) t_r + v_f (t_f-t_r) = \frac{1}{2}(v_i-v_f)t_r + v_f t_f.
\end{equation}
Setting this expression equal to Eq.(\ref{displ}) for coasting yields,
\begin{equation}
\label{tf}
t_f = \frac{(m/D) \ln\left(v_i/v_f\right)}{v_f} - \frac{v_i-v_f}{2v_f} t_r.
\end{equation}

\noindent This $t_f$ must be longer than the coasting time, $t_c$, of course.

Next, let us look at the battery SOC during the two stages - the initial regenerative-braking stage, followed by the constant-speed stage. First, how much energy can be transferred to the battery during the first stage?
Here kinetic energy is converted to electrical energy and then stored as chemical energy in the battery. Let us assume that the efficiency from kinetic to chemical-battery energy is given by $\eta$. We then have that 
\begin{equation}
\label{b1}
\Delta E_{batt}^{(1)} = \eta (|\Delta T|-W^{(1)}_{fric}),
\end{equation}
with $|\Delta T| = \frac{1}{2}m (v_i^2-v_f^2)$. Equation (\ref{b1}) can also be regarded as a definition of $\eta$, the regenerative efficiency \cite{verma, srbik}. But what is $W^{(1)}_{fric}$ for this stage?

We know that $W^{(1)}_{fric}=\int F_D dx = D\int_0^{t_r} v^3 dt$, where the speed is given by
\begin{equation}
v(t) = v_i - \left(\frac{v_i-v_f}{t_r}\right) t,
\end{equation}
describing the constant deceleration part shown in Fig.\ref{fig1} (blue line).  
If we denote $v_i-v_f$ by $w$, expand out the cubic term, and change variables of integration, we can show that,
\begin{equation}
\label{f1}
W^{(1)}_{fric} = D t_r (v_i^3-\frac{3}{2} v_i^2 w + v_i w^2 - \frac{1}{4} w^3)
\end{equation}
We can then substitute Eq.(\ref{f1}) into Eq.(\ref{b1}).

Now let's turn to the second stage. Here the electric motor has to simply counteract the drag force. The work done by the drag force is given by,
\begin{equation}
\label{W}
W^{(2)}_{fric} = D v_f^2 \Delta x_2 = D v_f^3 (t_f-t_r),
\end{equation}
with $t_f$ given by Eq.(\ref{tf}).

Assuming that the efficiency of converting chemical energy stored in the battery to electrical and then mechanical energy is $\epsilon$,
we arrive at,
\begin{equation}
\Delta E_{batt}^{(2)} = - \frac{1}{\epsilon} W^{(2)}_{fric},
\label{b2}
\end{equation}
where $W$ is given by Eq.(\ref{W}), and so $\Delta E_{batt} = \Delta E_{batt}^{(1)}+\Delta E_{batt}^{(2)}$. The problem is then reduced to determining the sign of $\Delta E_{batt}$. 

\begin{figure}
\includegraphics[width=0.4\textwidth]{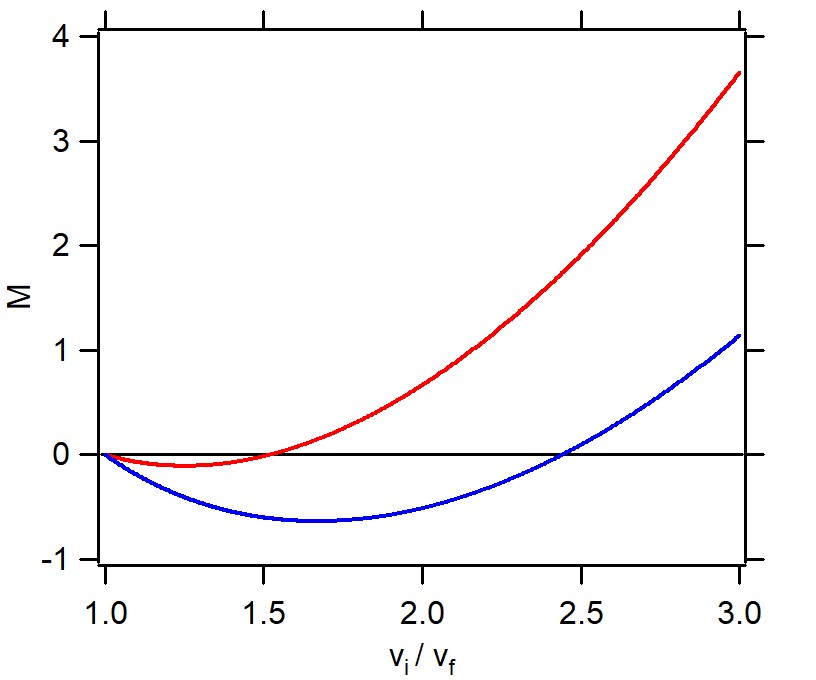}
\caption{The expression M of Eq.(\ref{M}) plotted as a function of the speed ratio. The red line is for $\epsilon=\eta=0.8$, and the blue line for $\epsilon=\eta=0.6$.}
\label{fig2}
\end{figure}

This expression for $\Delta E_{batt}$ is still fairly complicated and includes 7 parameters: $v_i$, $v_f$, $t_r$, $D$, $m$, $\eta$, and $\epsilon$. Let us simplify things by analyzing the limit of short $t_r$, such that $t_r << t_f$. In this limit the equations simplify significantly. We get,
\begin{equation}
\Delta E_{batt} = \frac{1}{2} m (v_i^2 - v_f^2) \eta - m v_f^2 \ln \left(\frac{v_i}{v_f}\right) \frac{1}{\epsilon}.
\end{equation}
In other words, we have to determine the sign of the following expression,
\begin{equation}
\label{M}
M = \left(\frac{v_i^2}{v_f^2}-1\right)\eta - 2 \ln\left(\frac{v_i}{v_f}\right)\frac{1}{\epsilon}. 
\end{equation}
Equation (\ref{M}) is plotted in Fig.\ref{fig2} as a function of speed ratio for two different sets of efficiencies. We see that for speed ratios only slightly larger than 1, $M$ is negative. For larger speed ratios, $M$ always turns positive. The crossing depends sensitively on the efficiencies of the motor and generator, with lower efficiency pushing that break-even point out to larger speed ratios.

The conclusion is that coasting is preferable when the final target speed is not substantially below the initial speed, i.e., for instances where the desired slowdown is only moderate. For instances, where a more dramatic slowdown is desired, aggressive regenerative braking wins out. A quick calculation reveals that if we want to slow down to half of our initial speed, say from 60 mph to 30 mph, and assuming that $\eta=\epsilon$, any efficiency larger than $\eta=\epsilon=$ 68 \% will favor regenerative braking.  

\subsection{A concrete example}
\label{example}
So far, the treatment has been theoretical. To make the conclusions more concrete, let us calculate actual numbers from a speed-down experiment with a Tesla (Model 3) \cite {model}. For concreteness, consider decreasing the speed from 50 mph (22.35 m/s) to 25 mph (11.18 m/s) and compare regenerative braking to coasting. The parameters of this vehicle are given as: the drag coefficient $C_d$=0.23, the frontal area $A$=2.22 m$^2$, the weight of the car (with a driver) is $m$=1280 kg \cite {parameter}. The density of dry air is $\rho$=1.225 kg/m$^3$.

Assuming an efficiency of $\epsilon=\eta=0.75$, $t_f$ is calculated from Eq.(\ref{tf}) as,
\begin{equation}
t_f=253.8-0.5 t_r.
\label{tfr}
\end{equation}
Next, from Eq.(\ref{b1}) and (\ref{b2}) and $\Delta E_{batt} = \Delta E_{batt}^{(1)}+\Delta E_{batt}^{(2)}$ we get,
\begin{equation}
\Delta E_{batt}=32103.7-354.7 t_r.
\label{dE}
\end{equation}
According to Eqs.~(\ref{tfr}) and (\ref{dE}), $t_f$ and $\Delta E_{batt}$ have a linear relationship with $t_r$, and %as shown in Fig.\ref{fig4}. 
$\Delta E_{batt}$ is positive only when $t_r<92$s. Furthermore, we can also get the time of coasting $t_c=183$ s from Eq. (\ref{tc}), and thus we explicitly verify the time-ordering, $t_r<t_c<t_f$ when $\Delta E_{batt}>0$.

%\begin{figure}
%\includegraphics[width=0.4\textwidth]{fig4a.jpg} 
%\includegraphics[width=0.4\textwidth]{fig4b.jpg}
%\caption{(a) $t_f$ versus $t_r$. (b) $\Delta E_{batt}$ versus $t_r$.}
%\label{fig4}
%\end{figure}
In an actual speed-down experiment with a Tesla Model 3 we were able to reach a minimum $t_r$ of around 5.0 s. This is much below the theoretical cutoff value of 92 s (where the $\Delta E_{batt}=0$), as calculated above. We have to remember, however, that the calculated $t_c$ and $t_r$ are too long compared to realistic numbers, as the real dissipation is not just in the form of air-drag, but also includes rolling friction (tire, road) and internal friction.   
Nonetheless, 5 s is significantly below both the 92 s, and it is also well below realistic coasting times. It is thus very likely that this speed-down experiment left the battery with a larger charge than what coasting would have accomplished. 

\section{Optimizing regenerative braking}
\label{optimal}
\subsection{Charging efficiency considerations}
From the previous discussion, and in particular from Eq.~(\ref{dE}), it becomes clear that without any further consideration of the charging efficiency involved, the best strategy is to brake very severely to reach the desired target speed as quickly as possible and then to remain at that constant speed for the rest of the way. This was also embodied in our examined limit of very short $t_r$. As we now show, a variational-calculus formulation of that problem confirms this observation. In an initial attempt to shift into this framework, we seek to maximize the following functional,
\begin{equation}
S=\int_0^T \mathscr{L} dt,
\label{action2}
\end{equation}
where
\begin{equation}
\mathscr{L}=\mathscr{L}[v,\dot{v}] = -m v \dot{v} - D v^3.
\label{action3}
\end{equation}
The expression for $\mathscr{L}$ is derived from the combination of $|dT|= - m v \dot{v} dt$ and $dW_{fric} = D v^3 dt$ (where $T$ denotes the kinetic energy).
The associated Euler-Lagrange equation,
\begin{equation}
\frac{d}{dt}\left(\frac{\partial \mathscr{L}}{\partial \dot{v}}\right) = \frac{\partial \mathscr{L}}{\partial v},
\label{EL}
\end{equation}
then quickly leads to the result that $v(t)=0$.

If we think about it a bit, this result should not be unexpected. Since we have assumed that the regenerative efficiency, $\eta$, is entirely constant, it makes sense that the energetically best thing we could do is to immediately slow down to zero. If we have a constraint that we must travel a certain distance, i.e., that $\int_0^T v(t) dt = d$, the analysis can be modified by adding a Lagrange multiplier, but the end result is similar: it is best to immediately slow down to a small constant speed (consistent with the distance, $d$).  
 
It is clear that the culprit for this trivial answer is our assumption of constant efficiency. In reality, what EVs do when they need to slow down quickly is to blend in the physical brakes to assist the regenerative braking. The more severe the required deceleration, the more heavily the EV relies on brake pads and rotors. This means that the efficiency of converting kinetic energy into battery energy goes down significantly as the braking action quickens. The battery can only accept so much power delivered to it, and this maximum power also depends sensitively on the battery SOC and temperature. Even before we reach that power limit, Ohmic losses tend to increase with charging power to reduce efficiency \cite{genovese, srbik}, and friction braking is often blended in \cite{eng3, eng4, verma}.

How should we model these effects on efficiency? To start, it is clear that the efficiency is a monotonically decreasing function of the braking power that the battery ideally would be expected to ``absorb''. This power, in turn, could be approximated by the reduction in kinetic energy of the vehicle, $P\approx -dT/dt = -mv\dot{v}$. Thus, we can write $\eta=\eta(p)=\eta(v,\dot{v})$. We will later choose a linearly decreasing function. Such choice of $\eta(P)$, while not accounting for the full complexity, does capture the overall effect on efficiency heuristically and keeps the mathematics tractable.  

\subsection{Optimal braking curves without distance constraints}
Informed by these considerations, we now seek to maximize the following functional:
\begin{equation}
S=\int_0^T \eta(v,\dot{v})\left(-mv\dot{v}-Dv^3\right)dt
\label{action}
\end{equation}
The function, $\eta(v,\dot{v})$, should be close to $\eta_0$ for low $v\dot{v}$ and then decrease for larger $v\dot{v}$. 

Let us next substitute this new $\mathscr{L}$ of Eq.(\ref{action}) into the Euler-Lagrange equation, Eq.(\ref{EL}), and this returns a fairly complicated equation with several terms:
\begin{eqnarray}
\label{govern1}
&\left[\frac{\partial \eta}{\partial v}-\frac{d}{dt}\left(\frac{\partial \eta}{\partial \dot{v}}\right)\right](-mv\dot{v}-Dv^3) = \\ \nonumber
&\eta(3Dv^2)-\dot{\eta} mv+\frac{\partial \eta}{\partial \dot{v}}(-mv\ddot{v}-m(\dot{v})^2-3Dv^2\dot{v}).
\end{eqnarray}
It is interesting to note that the first part of Eq.(\ref{govern1}) is in the form of an Euler-Lagrange equation involving $\eta$ as the functional.
We must now specify a functional form for $\eta$ to proceed. To keep things mathematically tractable, let us choose the following functional form for $\eta$:
\begin{equation}
\eta(v,\dot{v})=\eta_0+b(mv\dot{v}). 
\label{eta}
\end{equation}
Here we have assumed that the efficiency drops linearly with power; the two parameters are the y-intercept $\eta_0$, and the slope, $b$. Note that $b$ here is positive, since $\dot{v}$ is negative. A side benefit of this choice for $\eta$ is that it forces the left side of Eq.(\ref{govern1}) to zero, yielding:
\begin{equation}
\frac{\partial \eta}{\partial \dot{v}} (mv\ddot{v}+m(\dot{v})^2+3Dv^2\dot{v}) + mv \dot{\eta} - 3\eta D v^2  = 0.
\label{govern2}
\end{equation} 

\noindent From Eq.(\ref{eta}), $\frac{\partial \eta}{\partial \dot{v}}=bmv$ and $\frac{\partial \eta}{\partial v}=bm\dot{v}$, and we get,
\begin{equation}    
mv\dot{\eta}=mv\left(\frac{\partial \eta}{\partial v}\dot{v}+\frac{\partial \eta}{\partial \dot{v}}\ddot{v}\right) = bm^2v(\dot{v})^2+bm^2v^2\ddot{v}.
\label{subs}
\end{equation}
Substituting Eqs.(\ref{subs}) and (\ref{eta}) into Eq.(\ref{govern2}) yields (after a few steps) the following differential equation:
\begin{equation}
2bm^2v^2\ddot{v}+2bm^2v(\dot{v})^2- 3 \eta_0 D v^2 = 0.
\label{govern3a}
\end{equation}

Although not imperative, it can be helpful to non-dimensionalize Eq.(\ref{govern3a}). For this purpose we introduce the non-dimensional quantities $u$ and $\tau$, defined as $\tau=t/\alpha$, with $\alpha=\frac{m}{D v_i}$, and $u = \alpha (D/m) v = v/v_i$. Inserting these definitions into Eq.(\ref{govern3a}), we arrive at the governing equation:
\begin{equation}
u^2u''+u(u')^2-\gamma u^2=0.
\label{govern3b}
\end{equation}
Here the prime indicates differentiation with respect to $\tau$, and 
\begin{equation}
\gamma=\frac{3\alpha^3 \eta_0 D^2}{2 m^3 b} = \frac{3\eta_0}{2 D b v_i^3}.
\label{para}
\end{equation} 
Equation (\ref{govern3b}) represents a 2nd-order nonlinear differential equation that we can reduce to a first-order one. Dividing by $u^2$ and introducing $w=u'$, we can rewrite Eq.(\ref{govern3b}) in the following form,
\begin{equation}
\frac{dw}{du}+\frac{w}{u}=\frac{\gamma}{w}.
\end{equation}
We recognize this as a Bernoulli-type equation, which can be solved to obtain,
\begin{equation}
w(u)=-\frac{\sqrt{2\gamma u^3+k_1}}{\sqrt{3}u}.
\label{w}
\end{equation}
Here $k_1$ is an constant of integration that can be computed from the initial conditions as follows: since $w(1)=u'(\tau=0)=u'_0$, Eq.~(\ref{w}) leads to $k_1=3 (u'_0)^2-2\gamma$. 

To obtain $u(\tau)$, we remember that $u'=w$. Thus,
\begin{equation}
\int_1^u \frac{d\tilde{u}}{w(\tilde{u})} = \int_0^{\tau} d\tilde{\tau},
\end{equation}
and finally,
\begin{equation}
\tau = \int_{1}^u \frac{\sqrt{3} \tilde{u} d\tilde{u}}{\sqrt{2\gamma \tilde{u}^3+k_1}}.
\label{analyt}
\end{equation}
The integral on the right side of Eq.~(\ref{analyt}) can be evaluated in closed form (which would involve hypergeometric functions) or numerically integrated. %The final step then involves inverting $\tau(u)$ to obtain $u(\tau)$. 

Alternatively, we could of course solve the original differential equation, Eq.({\ref{govern3b}), numerically by specifying the initial conditions $(u(0),u'(0))$. In either case, it is important to estimate realistic values of $\gamma$ given by Eq.(\ref{para}). Every constant therein is straightforward with the exception of $b$. To get a reasonable estimate of this constant, we can start with maximum charging rates for the Tesla Model 3 (SR+) of about 100 kW. It is likely that for this vehicle, the maximum regenerative power is software-limited to the somewhat lower threshold of around 75 kW. In our mathematical model, of course, the efficiency goes down linearly with power. Let's assume that we reduce $\eta$ to $\eta_0/2$ for 75 kW. This then implies a value of $b=5*10^{-6}$s/J, which yields $\gamma=67.5$ for $\eta_0=0.75$ and $v_i=50$ mph.  

\begin{figure}
\includegraphics[width=0.35\textwidth]{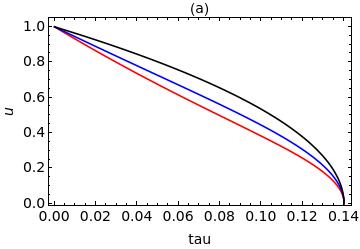} 
\includegraphics[width=0.35\textwidth]{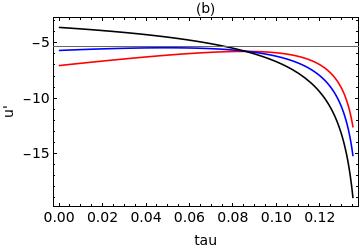}
\caption{Numerical solutions for $u(\tau)$ such that the stopping time is $0.14$. (a) Blue curve shows $\gamma=40$, red curve shows $\gamma=70$. For comparison, the black trace depicts the constant-power curve. (b) The same data plotted as an acceleration graph. All accelerations become extremely negative near $\tau=0.14$, so we plot only to 0.13 to highlight to essential differences between the traces.}
\label{fig3}
\end{figure} 
Figure \ref{fig3}(a) shows some typical numeric solutions obtained with {\it Mathematica} using the ``NDSolve'' command \cite{math}. For the blue trace we set $\gamma$ to 40, and for the red trace $\gamma=70$. The initial conditions were $u(0)=1$ for both, and $u'(0)$ was adjusted to obtain the same stopping time, $\tau_f=0.14$ (defined by $v(\tau_f)=0$). For the red trace, this implies $u'(0)=-7.01$, and for the blue, $u'(0)=-5.65$. It should be noted that we get indistinguishable curves by using Eq.~(\ref{analyt}) and inverting the resulting $\tau(u)$.

Also shown for comparison in Fig.\ref{fig3}(a) is the speed profile for a run where the extracted power is constant in time (black trace), with that constant adjusted to again yield the same stopping time. A straightforward calculation reveals that for constant power, $v(t)=\sqrt{v_0^2-\kappa t}$.

When comparing the red and black traces, it is evident that the optimal solution is one where initially, at high speeds, more power is extracted than for lower speeds. This makes sense, since the goal is to minimize losses from air-drag, and drawing out more energy from the available kinetic energy during the initial phase is advantageous. 

Notice also that, according to Eq.(\ref{para}), when the parameter $\gamma$ is raised, $b$ decreases, assuming the same EV is driven (identical $m, D, \eta_0$) and the same starting speed, $v_i$, is used. For smaller slope, $b$, governing $\eta(P)$, the regenerative braking can be made more severe without incurring additional efficiency penalties. This explains the more negative overall acceleration seen in the red trace at the start, when compared with the blue trace.
                                                                                                                                                                      
Finally, Fig.\ref{fig3}(b) plots the same data as an acceleration graph, $u'$ versus $\tau$. As before, the $\gamma=70$ solution appears in red, $\gamma=40$ in blue,  and the constant-power curve in black. For all curves, $u'$ diverges to negative infinity as $u\rightarrow 0$. However, it is also apparent that the optimal solutions for both values of $\gamma$ feature more negative accelerations at the beginning, and a less negative acceleration towards the end, compared to the constant-power curve. We again see that air-drag considerations nudge the optimal solutions in the direction of reducing speed at the beginning as much as possible. 

\begin{figure}
\includegraphics[width=0.35\textwidth]{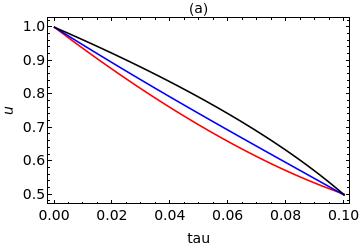}
\includegraphics[width=0.35\textwidth]{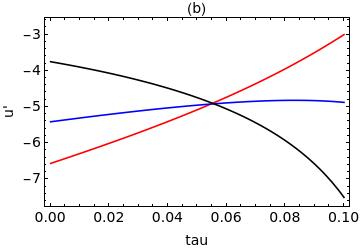}
\caption{Numerical solutions for $u(\tau)$ such that $u(0.1)=0.5$. (a) Blue curve shows $\gamma=40$, red curve shows $\gamma=70$. For comparison, the black trace depicts the constant-power curve. (b) The same data plotted as acceleration versus time. The differences between the various solutions are visually accentuated.}
\label{fig5}
\end{figure} 	
This feature of the optimal solution is also present when the vehicle does not come to a stop but reaches a non-zero final speed. Mathematically, we need only modify the endpoint condition to $v(T)=v_f$. Figure \ref{fig5}(a) depicts the three solutions of the previous figure but now for a final (reduced) speed of $u(\tau_f)=0.5$, with $\tau_f=0.1$. In Fig. \ref{fig5}(b), the same data is plotted as an acceleration graph that accentuates the differences and again showcases the strategy of making the acceleration as negative as possible at small times (red trace).

These optimized acceleration curves in Fig.\ref{fig5}(b) can be compared to accelerometer data from a real test-drive in a Tesla Model 3 where the speed was reduced from 50 mph to 25 mph. This measurement is shown in Fig.\ref{fig6}. We see that both the magnitude of acceleration, $|u'|$, and the recovered power are not constant but largest at short times. In that respect, the result resembles the red curve of Fig.\ref{fig5}(b). 
\begin{figure}
\includegraphics[width=0.4\textwidth]{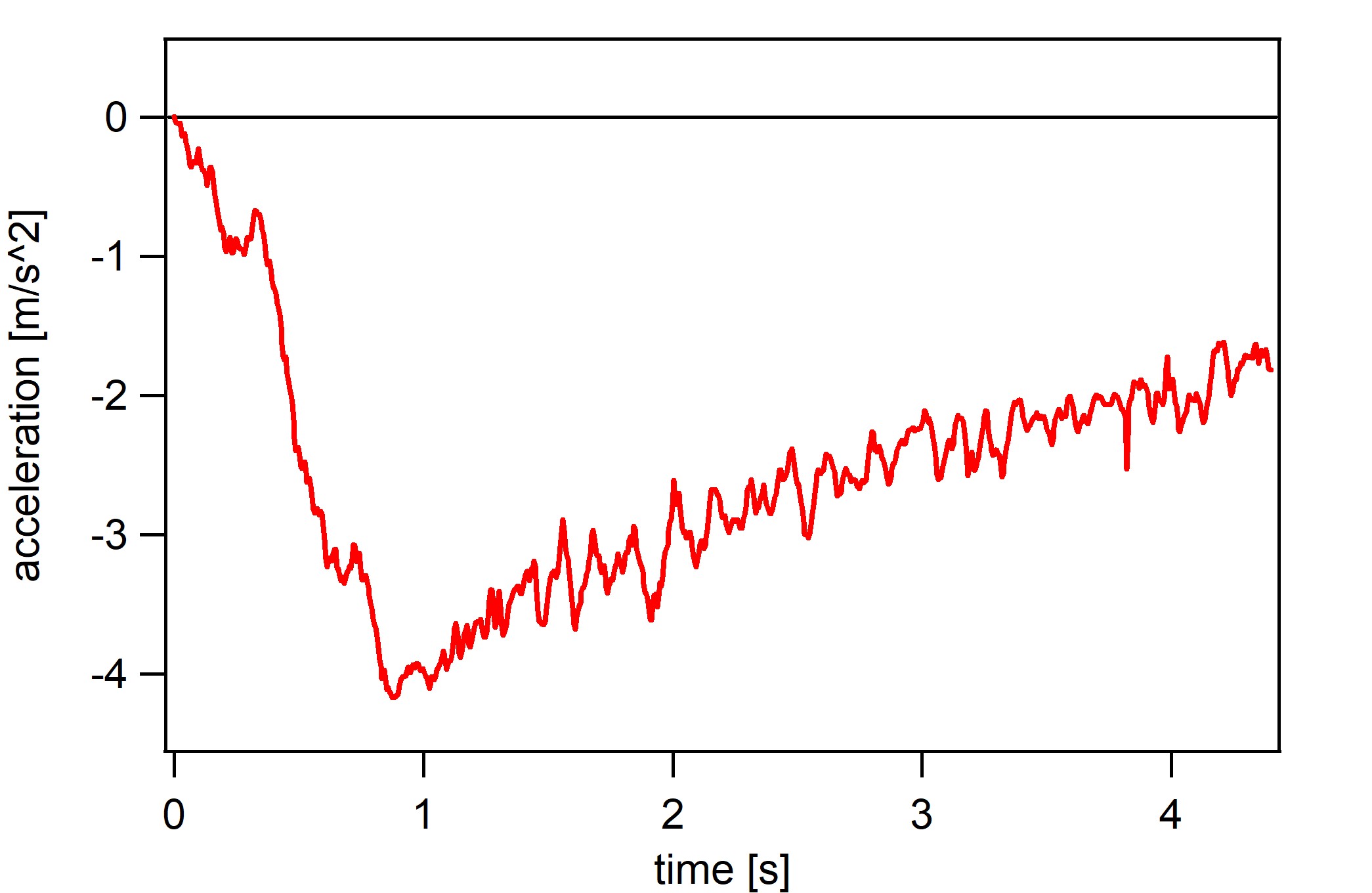}
\caption{Accelerometer data from a test drive in a Tesla Model 3 (SR+) where the speed was reduced from 50 mph to 25 mph using full regenerative braking. We see that the acceleration is indeed largest initially when the speed is close to 50 mph.}
\label{fig6}
\end{figure} 

We conclude this section by circling back to the original question: how much energy is added or drawn from the battery after performing braking actions according to these optimized solutions?  $\Delta E_{batt}$ is, in fact, given by the functional of Eq.(\ref{action}). Should this integral, when evaluated for the numerical solutions found earlier, be positive, the battery would ``gain charge'' during the examined time interval. Casting this definite integral in terms of the non-dimensional variables, $u$ and $\tau$, we get that $\eta(u,u')=\eta_0 \left(1+\frac{3}{2\gamma} u u'\right)$, and:
\begin{equation}
\Delta E_{batt} = \eta_0 (mv_i^2) \int_0^{\tau_f} \left(1+\frac{3}{2\gamma} u u'\right) \left(-u u'-u^3 \right) d\tau.
\end{equation} 
We use the ``NIntegrate'' command in {\it Mathematica} to evaluate this integral for the traces in Fig.\ref{fig3} (where $\tau_f=0.14$). We thus obtain 0.315 $m v_i^2$ for the red trace, 0.293 $m v_i^2$ for the blue trace. Both integrals are therefore positive and less than $\eta_0 (T_i-T_f)$. Both solutions are also improvements over the constant-power curve; when compared to the red trace, it recovers roughly 2\% less energy.

Similarly, we can compare the optimal braking curves in Fig.~\ref{fig5} to a the constant acceleration case considered earlier in Sec.\ref{sec1}, where the speed drops linearly with time. When we do this for the $\gamma=70$ curve (red trace), for instance, we find that the optimal solution recovers 0.226 $mv_i^2$, which is only about 0.1 \% more energy than constant-acceleration braking. In a sense, this observation retroactively validates the initial, more simplified approach.    

\subsection{Adding a distance constraint with a Lagrange multiplier}
We can ask a slightly modified question: what is the optimal braking curve, $v(t)$, that connects the two points, $v(0)=v_i$ and $v(T)=v_f$, given that the car must cover a specified distance? Notice that we now have added another condition at the end. We would like to consider curves of equal displacment, $d=\int_0^{T} v dt$.
Variational calculus tells us that we now need to make the following integral stationary:
\begin{equation}
\int_0^{T} \left(\mathscr{L} + \lambda v\right) dt,
\end{equation}  
where $\lambda$ is called the Lagrange multiplier and is related to the displacement.
The Euler-Lagrange equation now reads,
\begin{equation}
\frac{d}{dt} \left(\frac{\partial \mathscr{L}}{\partial \dot{v}}\right) = \left(\frac{\partial \mathscr{L}}{\partial v}\right) + \lambda.
\end{equation}
Following similar steps as before, we thus arrive at the modified non-dimensional governing equation,
\begin{equation}
u^2u''+u(u')^2-\gamma u^2 - \kappa \lambda = 0,
\label{lam}
\end{equation}
with $\kappa=\frac{\gamma}{3\eta_0 D v_i^2}$ such that $\kappa \lambda$ is non-dimensional.

Employing the same analytical technique of reducing Eq.~(\ref{lam}) to a first-order equation yields,
\begin{equation}
\tau = \int_{1}^u \frac{\sqrt{3} \tilde{u} d\tilde{u}}{\sqrt{2\gamma \tilde{u}^3+6 \kappa \lambda + k}},
\end{equation} 
with $k = 3(u_0')^2 - 2\gamma - 6 \kappa \lambda$.

Figure \ref{fig7} shows the effect of the distance constraint. The three traces correspond to three different values of the Lagrange multiplier and thus braking distances, with $\gamma$ set to 70. We know that $d=\int v dt = \frac{m}{D} \int u d\tau$. Therefore, to find the actual distance traveled, we numerically integrate the traces in Fig. \ref{fig7}. Then, using the values in section \ref{example}, we obtain $d=307, 295,$ and $282$ m, respectively for the red, black and blue traces. Furthermore, $T=\alpha \tau_f=\alpha (0.1)=18.3$ s. The recovered energies for the red, black and blue trace, respectively, are 0.2243 $mv_i^2$, 0.2260 $mv_i^2$, and 0.2245 $mv_i^2$. Not surprisingly, the optimal solution without displacement constraint (black trace) outperforms the other two solutions with the Lagrange multipliers. 

\begin{figure}[h]
\includegraphics[width=0.375\textwidth]{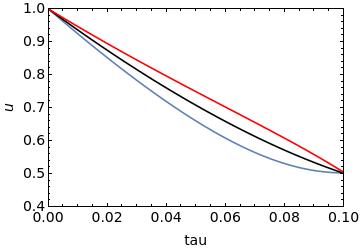}
\caption{The optimal braking curves for three different values of the Lagrange multiplier $\kappa \lambda = \lambda^*$. The black, red and blue curves shows the results for $\lambda^*=$0, 19.2, and -19.2, respectively, and $\gamma=70$.}
\label{fig7}
\end{figure}

\section{Conclusion}
We can now answer the original question posed in the title with confidence. Under many circumstances, it is advantageous to employ aggressive regenerative braking over coasting. We have explored this from a number of different angles, choosing initially a simplified assumption of constant braking acceleration. In the second part of the paper, we then found more realistic braking curves - ones that maximize battery charge, after making certain assumptions on the efficiency profile for regenerative battery charging. To keep the problem mathematically tractable, we used a linear efficiency model, which allowed us to formulate the associated Euler-Lagrange equation. We then found both numerical and analytical (integral-form) solutions to the Euler-Lagrange equation. Finally, we explored the effect of introducing the constraint of fixed displacement (distance traveled) via a Lagrange multiplier. Not unexpectedly, one lesson from the optimal braking curves obtained in this manner is that it pays to make the acceleration as negative as possible at the beginning, in an effort to reduce air-drag losses as much as possible.

\end{document}